# Research on the Software of Track Reconstruction in Vertex Chamber of BESII[*]

ZENG Yu[1)]    MAO Ze-Pu

(Institute of High Energy Physics, Chinese Academy of Sciences, Beijing 100039, P.R.China)

**Abstract**   The software of track reconstruction of the vertex chamber of BESII-VCJULI was studied when it was transplanted from the HP-Unix platform to PC-Linux. The problems of distinct dictionary storage and precision treatment in these two different platforms were found and settled in the modified software. Then the obvious differences of the candidate track number in an event and some track parameters caused by them were reduced from 74% to 0.02% and from 5% to 0.5%, respectively. As a result, the quality of the track finding was greatly improved and the CPU time saved.

## 1   Introduction

The Beijing Spectrometer II (BESII), which records information from electron positron interactions in the tau-charm energy region from 3-5GeV to study $\tau$ leptons, D mesons, J/$\psi$, and $\psi'$ physics, consists of several components [1]: the vertex chamber (VC), the main drift chamber (MDC), the time of flight counters (TOF), the shower counters (SC), muon counters ($\mu$C), luminosity monitor (LUM), solenoid magnet, electronics and its calibration system.

The BESII vertex chamber consists of 640 aluminized Mylar drift tubes distributed in totally twelve layers. Except for the four middle stereo layers, which are used in Z directional track finding, the remaining eight layers are all axial ones used in track finding in r-$\phi$ plane. The whole twelve layers of tubers are divided into six snake layers, each snake layer is composed of two adjacent wire layers from innermost to outermost. The two wire layers in one snake layer is offset by 1/2 tube to ensure that each will have at least two high resolution measurements in each group of four wire layers.

The vertex chamber can be used as a trigger to suppress background, and it can be also used in matching reconstruction with other sub-detectors. The study of matching reconstruction of MDC and VC shows the resolution of momentum is greatly improved when VC is used in track reconstruction of BESII, and 10MeV is improved when analyze J/$\psi \to$e+e– [2].

Before 2000, the environment of reconstruction of BES is made up of HP-Unix system, and all the reconstruction software can run on HP-Unix platform. With the rapid development of PC machines, they have considerably superior cost performance to HP ones, however. IHEP thus began to adopt PC-FARM to construct its environment of physics analysis [3] just as other HEP labs do. PC-FARM is a multi-PC computation environment with many characteristics such as powerful computation ability, good cost performance, convenient management and handy upgrade etc., which make it an ideal program to establish large-scale computation environment. Because of the difference between PC-Linux and HP-Unix, hardware and software included, the reconstruction results of BES events on two platforms have obvious differences. To make full use of the superiorities of PC machines, systematic study of reconstruction software must be done when the software is transplanted from HP-Unix platform to PC-Linux platform.

---

[*] High Energy Phys. and Nucl. Phys., 2003, 27(9): 808-812
∗ Supported by National Natural Science Foundation of China (19991480)
[1)] E-mail: zengy@mail.ihep.ac.cn

## 2 Track reconstruction

The function of track reconstruction is to redisplay particle's track of flight by making use of raw data. According to the principal that the projection of a particle's track is an arc when the charged particle flies in the magnetic field, values such as the momentum, the charge and so on are obtained.

The software of track reconstruction in vertex chamber (VCJULI [4]) consists of two parts: track finding and track fitting.

1) By the use of track dictionary, the part of track finding, adopting the method of global track recognition [5], has the characteristic of rapidness and high efficiency. Track dictionary consists of template dictionary, cell dictionary and auxiliary dictionary. Among them, all the constituents of cells are stored in template dictionary and all the track patterns can be found by using MINUIT [6]; Cell dictionary is a list of templates, which pass through each cell in the chamber; Auxiliary dictionary is generated to suppress spurious tracks [7]. All the work is done in r-$\phi$ plane with the information provided by axial layers. When taking into account the information such as time of drift, geometric position and the principal that the projection of a charged particle's track is an arc, related parameters can be obtained through fitting [7].

2) Parameters obtained in r-$\phi$ plane and information supplied by stereo layers are used in track fitting to get Z position. According to the function of charged particle's movement in even magnetic field [7], five parameters describing the helix function, i.e., the position and the momentum of particle [8] are thus obtained.

## 3 The study of reconstruction software

With the effort of many people of BES collaboration in transplanting software from HP-Unix platform to PC-Linux platform, BES offline analysis software (Version 102) has successfully been used in the analysis of J/$\psi$ data. Track number and track parameters have distinguished differences between two platforms, however. Systematic check and study show the main reasons as listed in the following:

1) Although Linux system and Unix system are very much alike, the FORTRAN compilation of the BES code on Linux requires strict adherence to FORTRAN rules, however, it is less stricter on Unix [9];

2) There still exists in the BES code, originated from old FORTRAN version 18 years ago, many irregular codes, which lead to different results on those two platforms or cannot even run on Linux.

3) Current PC machines use 32 bits in calculation while HP workstations use 64 bits. Differences thus occur when some high precision in calculations are carried out on those two systems.

4) Different data storage and transaction exist between two platforms.

Therefore, to establish the environment of the BES offline analysis on PC-Linux platform, systematic check and study of the BES code on HP-Unix are required. To reconstruct on PC platform, we have studied the BES reconstruction software (V102) in detail.

The procedure of check can be divided into two steps: ①Reconstruct on two platforms respectively, then gather statistics of track numbers of each event and compare the track numbers of each subtype; ②Compare the parameters of each track when the track numbers found on HP-Unix and PC-Linux have no difference. We define that if the difference is greater than $1\% \times (R_{iU} - R_{iL})/R_{iU}$ ($R_{iU}$ and $R_{iL}$ are the values of '$i$'th parameter on HP-Unix and PC-Linux, respectively), the parameters on two platforms are different. The study of VC shows track number and track parameter on HP-Unix and PC-Linux are quite different, the main reason is distinct dictionary storage and precision treatment in these two different platforms.

## 3.1 check on track number

Through reconstruction on data Run15254.raw, we found 74% events have different track numbers. The main reason is:

The dictionary of MDC is generated in the type of binary system, which will not cause differences in two platforms. The dictionary of VC is generated by means of ASCII, however. Since the storage of an integer on those two platforms are different and 'bit operation' is required in track finding, ASCII number is read in from the dictionary and the value is transferred through the equivalence with a logical variable. However, when execute the equivalence between a logical variable and an integer one, the way to read and store is significantly distinct, which results in the differences in candidate track number.

Considering the equivalence of one-byte array (each cell occupies 1 byte) VCSDCT(100000) and four-byte array (each cell occupies 4 bytes) VCSDC4(25000)(that is, EQUIVALENCE (VCSDC4(1),VCSDCT(1))), e.g., for value $2047=(11111111111)_2$, the way to deal with is shown in Figure 1.

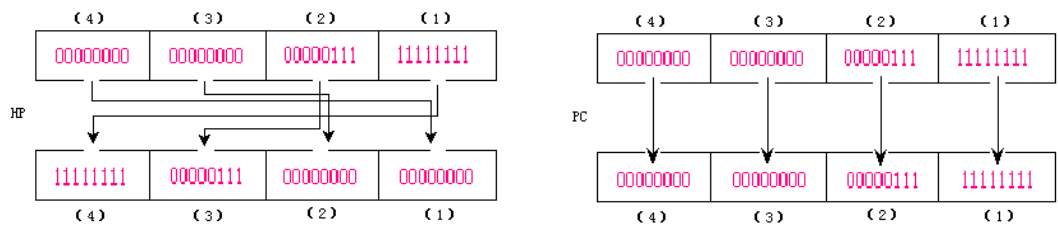

Figure 1.  One way to deal with 'dictionary' on HP-Unix and PC-Linux, respectively

As can be seen, if we need to get value from a certain byte, for example, from byte(2), the result on HP-Unix is $(00000000)_2$, however $(00000111)_2$ on PC-Linux. When converted into decimalism, it is 0 on HP-Unix and 7 on PC-Linux. Through comparison, we found the storage of each four byte is totally reversed on HP platform.

By the same token, as to the equivalence of one-byte array (each cell occupies one byte) VCSDCT(100000) and two-byte array (each cell occupies two bytes) (i.e., EQUIVALENCE(VCSDC2(1),VCSDCT(1))), considering the same value $2047=(11111111111)_2$, to get value from byte(2), for example. It is $(11111111)_2$, equals to –1 when converted into decimalism on HP platform while $(00000111)_2$ , equals to 7 when converted into the decimal system on PC platform. See Figure 2.

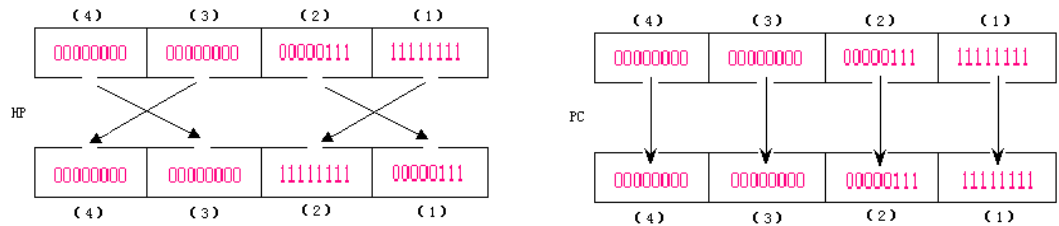

**Figure 2.** Another way to deal with 'dictionary' on HP-Unix and PC-Linux

As can be seen from the above, we can find the differences and the law in equivalence between one-byte array and four-byte array and the law in equivalence between one-byte array and two-byte array. Because of these reasons between HP platform and PC platform, program VCTKFD.f (used to search candidate track) get different value from array VCSDCT(*), where the contents of dictionary are stored. Thus, candidate track number found in r-ϕ turns out to be different. This problem is successfully solved through the settlement of the software.

With data Run15254.raw, the result of reconstruction on 5,000 events is shown in Figure 3. No.1 and No.13 on horizontal coordinate represent candidate track number and final track number of VC, respectively, while vertical coordinate represents the corresponding ratio of the events that have differences in track number in two platforms. Among them, candidate track is the track found in r-ϕ plane, and final track can be obtained when information of stereo layers are added.

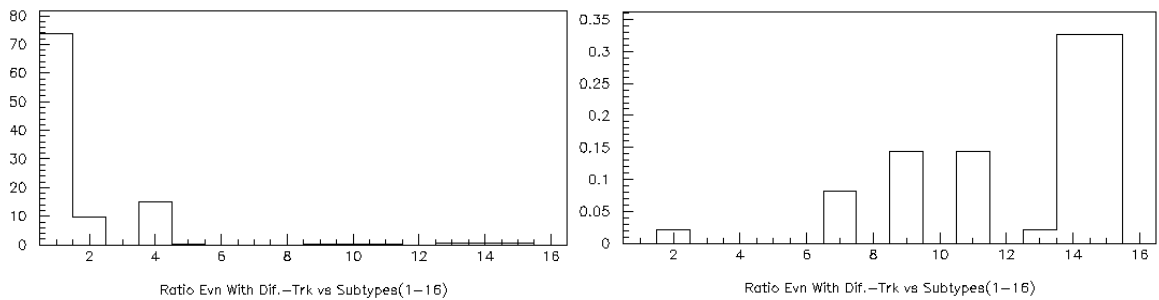

**Figure 3. Ratio of events with different candidate track number before modification and after modification**

1 VC candidate track；2 Neutral events；3 MDC;4 TOF;5 SC;6 Muon;7 dE/dx;8 Vertex Constrained Track Fit;

9 MDC Pointers;10 SC Pointers;11 Projected/Decay Coordinates of MDC Tracks;12 Virtual Track;

13 VC;14 VC Pointers;15 Projected/Decay Coordinates of VC Tracks;16 MDC-VC Matching Info.

As can be seen from Figure 3., the number of events with different candidate track number in r-ϕ plane has reduced from 3611 to 1 (event ID: 828), that is, from 74% (before modification) to 0.2‰ (after modification). Meanwhile, although the way to deal with the dictionary in PC platform and HP platform is different, which results in the distinction of candidate track number, final track number has little difference. The main reason is: The software will do the same 'bit operation' on most content of the dictionary, thus, taking into account the first 'bit operation' in track finding in r-ϕ plane, the software has operated even times which makes little differences in final track number on two platforms, however, track finding efficiency is very low. As to the part has not been carried out with the same 'bit operation' when adding information of stereo layers, that is, only been operated odd times, final track number will differ. Besides, after modification on the software of VC, other parts related to VC, such as TOF, SC (No4 and No5 in horizontal coordinate showed in Figure 3.) will have corresponding variation. Careful study of the event (event ID 828) shows that if run the event alone, there is no difference between two platforms. Difference occurs when run a series of events passing through it, however. The reason is the software is not strict enough and some rules on HP-Unix is somewhat not rigid, consequently, track number on HP-Unix is one more that on PC-Linux. See Figure 4.

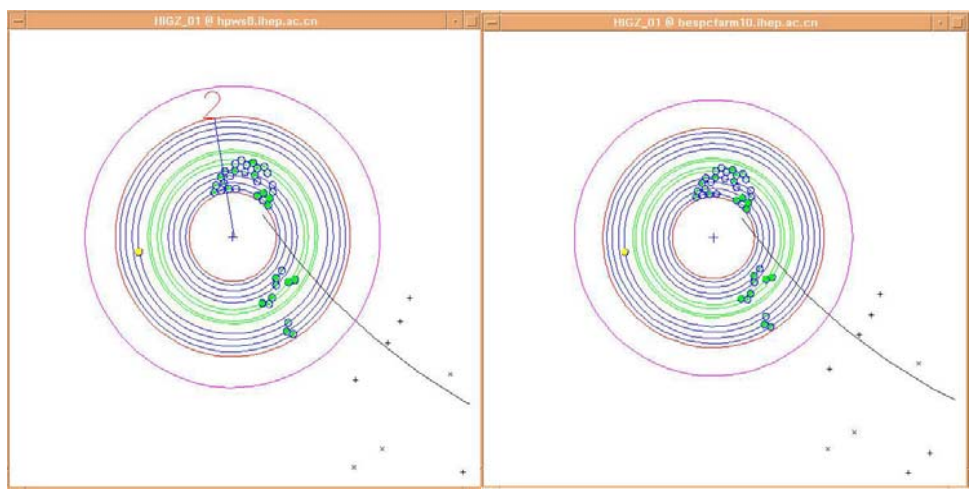

Figure 4. Display of event 828 on HP-Unix and PC-Linux, respectively

Therefore, after modification on the software in charge of reading the dictionary, candidate track number is completely the same and so is final track number.

### 3.2 check on parameters

To improve the precision of physics analysis and reduce the differences of analysis results between different platforms, each parameter of an event is required to be identical.

The reason why track parameter differs is determined by the following reasons: PC machines calculate in 32 bits while HP workstation in 64 bits. Differences occur when deal with some high precision calculations. After some modifications, including the change of variable type from single precision to double one, the differences in track parameter decreased significantly. Figure 5. shows the results of reconstruction on 5,000 events with data Run15254.raw on two platforms.

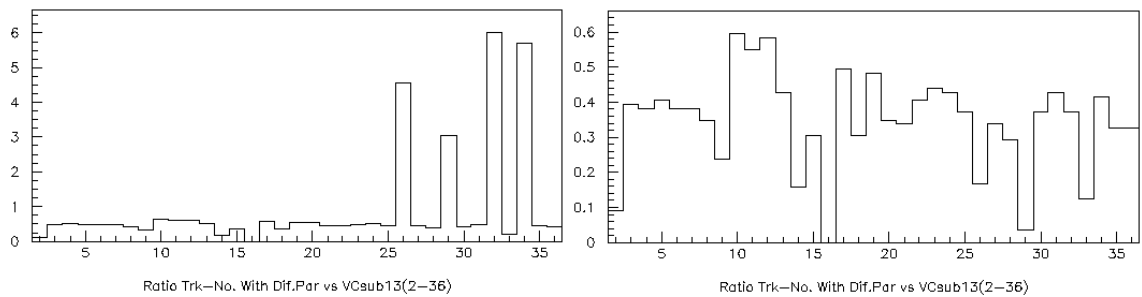

**Figure 5. Ratio of events with different track parameter before modification and after modification**

As shown in Figure 5., parameter 26,29,32,34 (parameters related to error matrix) has decreased from about 5% to less than 5‰. Ratio of events on some important track parameters, such as $P_x, P_y, P_z, P, \theta, \phi$ (4-9), is less than 4%.

## 4 Summary

After systematical check and modification on the track reconstruction software on Vertex Chamber of BESII on PC-Linux, the quality and efficiency of track finding have been greatly improved. Candidate track number has become completely the same on two platforms, and the ratio of events with different track parameter has reduced to 5‰ or so. Since VC locates at inner part of spectrometer and matches with TOF, MDC and ESC in track extrapolation, to further improve the precision in track extrapolation, check and modification are required in corresponding programs.

We would like to thank the staffs of Calibration Group at the Institute of High Energy Physics, Beijing. This work was supported by the National Natural Science Foundation of China under Contact No.19991480.